\documentclass[a4paper,aps,pra,twocolumn,
showpacs,preprintnumbers]{revtex4}
\usepackage[dvips]{graphicx}
\usepackage[centertags]{amsmath}
\usepackage{amsfonts}
\usepackage{amssymb}

\begin{document}

\title{Non-Linear Dynamics of Continuously Measured Bose-Einstein Condensates in One-Dimensional Harmonic Traps}

\author{T.Yu.~Ivanova}
\address{St.~Petersburg State University,\\
Ul'yanovskaya 5, 198504 Petrodvoretz, St.~Petersburg, Russia}

\author{M.S.~Samoylova}
\address{St.~Petersburg State University,\\
Ul'yanovskaya 5, 198504 Petrodvoretz, St.~Petersburg, Russia}

\author{D.A.~Ivanov}
\address{St.~Petersburg State University,\\
Ul'yanovskaya 5, 198504 Petrodvoretz, St.~Petersburg, Russia}

\date{\today}

\begin{abstract}
Continuous center-of-mass position measurements performed on an
interacting harmonically trapped Bose-gas are considered.  Using
both semi-analytical mean-field approach and completely quantum
numerical technique based on positive P-representation, it is
demonstrated that the atomic delocalization due to the measurement
back action is smaller for a strongly interacting gas. The
numerically calculated second-order correlation functions
demonstrate appearance of atomic bunching as a result of the
center-of-mass measurement. Though being rather small the bunching
is present also for strongly interacting gas which is in contrast
with the case of unperturbed gas. The performed analysis allows to
speculate that for relatively strong interactions the size of atomic
cloud determined with a single snapshot measurement can become
smaller than the ground-state cloud size.

\end{abstract}


\maketitle

\section{Introduction}
    \label{sec:intro}
Now that generation of the trapped degenerate gases has become
almost routine procedure in many laboratories the experimental and
theoretical activities are directed towards applications of these
systems. New frequency standards~\cite{atomic-clock} and various
types of weak-force detectors~\cite{drummond-leuchs,casimir-polder}
belong to such applications. Another important application of such
quantum systems is testing quantum-mechanical predictions in
mesoscopic regimes.

All these experiments involve measurements of some property of the
trapped gas. As is extensively discussed in quantum theory, the
measurements of small objects do inevitably disturb them even if
they are prepared as so called quantum non-demolition~\cite{QND}
ones. This issue becomes especially important if the measurement is
continuously performed during an experiment. Recently a few
experiments have been reported where the conditions required for
continuous BEC measurement have been
realized~\cite{brennecke,colombe,gupta,murch}. One of these works,
Ref.~\cite{murch}, especially addresses the measurement back-action
on the trapped non-interacting gas. Detailed understanding of
measurement back-action is very important for successful realization
of quantum-limited feedback control of the trapped
gas~\cite{wiseman-bec,wallentowitz-prl}.

An important degree of freedom is the collective or center-of-mass
position of the trapped gas. This degree of freedom can be accessed
via bringing the system into interaction with a few-mode external
field. Hence, this type of measurements is conceptually simpler than
a measurement resolving internal structure of the trapped cloud.
Moreover, the collective degree of freedom is proven to be
experimentally accessible in recent works cited
above~\cite{brennecke,colombe,gupta,murch}. These facts motivate the
theoretical analysis of quantum motion of a degenerate gas with the
continuously measured center-of-mass position that we present in
this article.

Unlike some other
works~\cite{drummond-leuchs,murch,wallentowitz-prl}, where similar
problems are considered, we focus on a gas of \textit{interacting}
particles. It is known that the center-of-mass (CM) motion of a
harmonically trapped gas is not coupled to the relative degrees of
freedom~\cite{pitaevskii-rmp}. This means that the inter-particle
interactions do not affect the quantum dynamics of the center of
mass. The CM measurement does, however, influence the properties
that depend on the relative motion such as particle density
distribution~\cite{wallentowitz-prl}. These properties are also
affected by the inter-particle interaction. Thus an interesting
question arises: does the simultaneous action of the CM position
measurement and the interaction result in some nontrivial dynamics
of the gas?  In other words, how does the interaction strength
influence the behavior of the measured gas?  It should be mentioned
that interesting effects in a different system that also contains an
open interacting BEC have recently been discussed in Ref.~\cite{ng}.

Considering a harmonically trapped repulsively interacting 1D Bose
gas we show that interaction-induced nonlinearity provides a
mechanism that partially compensates measurement back action and
stabilizes the cloud spreading. This can be accompanied by the
narrowing of the instant density profile of the trapped gas compared
with the ground-state density profile. Note that the master equation
used in this article to describe the effect of the measurement also
describes situations where the CM of the gas is weakly coupled to a
hot reservoir~\cite{petrucione-book}. To some extent such coupling
might describe quantum fluctuations of the trapping potential that
are always present in an experiment. Thus the discussed effects
might be relevant even for the cases where there is no explicit
observation or control.

We base our discussions on the numerical solution of the many-body
quantum problem using positive P-representation of the density
operator~\cite{drummond-gardiner}. Although this approach is known
to be limited to relatively small evolution
times~\cite{boundary-terms} it seems to be least resource consuming
allowing to obtain reasonable results on a single desktop computer.
As a complement to numerical results we derive simplified mean-field
approach that qualitatively agrees with our numerical results.

The article is organized as follows. In Sec.~\ref{sec:noise} the
model that is considered is described and examples of possible
physical implementations of the CM position measurements are given.
The mean-field approach is derived and discussed in
Sec.~\ref{sec:single-atom}. Section~\ref{sec:numeric} is devoted to
the numerical analysis of the system and to the comparison of the
results with the mean-field predictions. In Sec.~\ref{sec:summary}
we present the summary of the results of the article.

\section{Model}
    \label{sec:noise}

Let us consider a gas of spin-0 bosons with mass $m$ interacting via
a repulsive $\delta$-function potential. The gas is trapped in a
cylindrically symmetric potential along $x$ axis with frequencies
$\omega_0$ and $\omega_{\bot}$ in axial and radial directions,
respectively; $a_0 \!=\! (m\omega_0)^{-1/2}$ and $a_{\bot} \!=\!
(m\omega_{\bot})^{-1/2}$~\footnote{Hereinafter we use $\hbar \!=\!
1$.} are the corresponding lengths of the harmonic oscillator ground
state. In the case of tight confinement in radial direction
($\omega_0 \ll \omega_{\bot}$) and at low enough temperature all
atoms are in the ground state of the radial potential and the gas
can be considered as being effectively a trapped $1$D gas. In
quasi-$1$D case given that $a_{\bot} \gg a_0$ the interaction of
bosons can be described by the coupling constant $g_{1D} \!=\! 2 a_s
\omega_{\bot}$~\cite{coupling-const}, where $a_s$ is the s-wave
scattering length.

We assume that the center-of-mass or, more precisely, collective
position of the gas in a trap is continuously observed. Strictly
speaking one should distinguish between the center-of-mass and
collective position~\cite{drummond-leuchs}. The former is the true
mean value that accounts for particle number fluctuations while the
latter is merely the weighted sum of particles coordinates. Below we
will deal with states having small particle number fluctuations and
neglect the mentioned difference.

As a physical model of such a measurement one can suggest, for
example, the approach discussed in~\cite{mekhov} and experimentally
realized in~\cite{colombe,murch}. In these experiment the
$\mbox{}^{87}\mathrm{Rb}$ atoms trapped in an atom chip have been
coupled to high-finesse optical cavity. The authors show that the
atoms can be localized entirely within a single antinode of the
standing-wave cavity field. Under these conditions the collective
atomic position determines the atom-field coupling strength and can
be determined by measuring the cavity transmission.

Another straightforward example is Faraday rotation of the polarization of
non-resonant light passing through the gas placed in a
non-homogenous magnetic field. Within the framework of the classical
theory~\cite{hutchinson} one can easily find the following
expression for the total rotation angle
\begin{equation}\label{eq:Faraday}
    \phi = \frac{2 \pi e^3 \omega^2}{m_e^2 c^2 (\omega_{\rm R}^2 - \omega^2)^2} \int
    n(x) B(x) dx,
\end{equation}
where $e$ is the charge of an electron, $m_e$ is the mass of an
electron, $c$ is the velocity of light in a vacuum, $\omega$ is the
probe light frequency, $\omega_{\rm R}$ is the atomic resonance
frequency, $n(x)$ is the atom density, $B(x)$ is the magnetic field.
Here we integrate over the atomic sample thickness and neglect a
deviation of the refractive index from unity. It is seen from
Eq.~(\ref{eq:Faraday}) that in the case of the linear magnetic field
the Faraday rotation is proportional to the center-of-mass
coordinate of the atoms $\bar{X} \!=\! (1/N)\int dx x n(x)$, where
$N$ is the estimate of the number of atoms that, for simplicity,
might be thought of as the average number of atoms. One certainly
can think of other schemes to measure the CM coordinate. Therefore,
below we do not concentrate on a particular realization but consider
the effects typical for all possible experimental approaches to CM
position measurements.

Our main concern is the measurement back action, thus we ignore
particular measurement outcomes and analyze unconditioned dynamics.
In this case the system is described by the master equation
\begin{equation}
    \label{eq:master}
    \dot{\hat{\rho}} =
    -i[\hat{H},\hat{\rho}]-\kappa[\hat{X},[\hat{X},\hat{\rho}]].
\end{equation}
The first term on the right-hand side of this equation corresponds
to the hamiltonian evolution. In second quantization the Hamiltonian
has the form
\begin{eqnarray}
    \label{eq:ham}
    \hat{H} &=& \int  \Big[-\frac{1}{2m} \hat{\Phi}^{\dag}(x)\partial_x^2
    \hat{\Phi}(x) + \frac{m \omega_0^2}{2} x^2 \hat{\Phi}^{\dag}(x)\hat{\Phi}(x)
    \nonumber \\
    &+& g_{\rm 1D} \hat{\Phi}^{\dag}(x)^2\hat{\Phi}(x)^2 \Big] dx,
\end{eqnarray}
where $\hat{\Phi}(x)$ and $\hat{\Phi}^{\dag}(x)$ are bosonic field
operators obeying the commutation relation
$[\hat{\Phi}(x),\hat{\Phi}^{\dag}(x')] \!=\! \delta(x \!-\! x')$.
The second term in Eq.~(\ref{eq:master}) describes the measurement
of the CM coordinate of the atoms
\begin{equation}
    \label{eq:col-coord}
    \hat{X} = \frac{1}{N} \int x \hat{\Phi}^{\dag}(x)\hat{\Phi}(x) dx.
\end{equation}
The parameter $\kappa$ characterizes the measurement resolution of
the apparatus and determines the back-action strength. Note that the
same master equation can be obtained if all the atoms are weakly
coupled to the same heat bath of high temperature. Namely, taking
these limits into account Eq.~(\ref{eq:master}) follows from the
well known Caldera-Legget master equation~\cite{petrucione-book}.

\section{Mean-field approximation} \label{sec:single-atom}

The evolution of the continuously measured interacting Bose gas can
be found from the Hamiltonian~(\ref{eq:ham}) and the master
equation~(\ref{eq:master}). However, even for the moderate numbers
of atoms $N$ the direct numerical integration of the master equation
is impracticable due to the large dimensionality of the $N$-atom
Hilbert space. This principle problem of many-particle physics can
be attacked with various numerical methods. In this article we use
approach based on positive P-representation of the density
operator~\cite{drummond-gardiner}. However, before presenting
numerical results we give a simple mean-field consideration that
qualitatively describes the system behavior. It is shown that even
this approach reveals some interesting features in the dynamics of
the continuously measured interacting Bose gas.

We start by defining a single-atom density matrix as
\begin{equation}
    \rho(x_1,x_2) = \mathrm{Tr}\{\hat{\Phi}^{\dag}(x_1)
    \hat{\Phi}(x_2)\hat{\rho}\}.
\end{equation}
The evolution of the single-atom density matrix is described by the
following equation
\begin{eqnarray}
\label{eq:density-matrix}
    \dot{\rho}(x_1,x_2) &=& \Big
    \{i\frac{m}{2} (\partial_{x_2}^2 - \partial_{x_1}^2) -
    i \frac{m \omega_0^2}{2}(x_2^2 - x_1^2)\nonumber \\
    &+& 2i g_{\rm 1D}[n(x_1)-n(x_2)] \nonumber \\
    &-& \frac{\kappa}{N^2} (x_2-x_1)^2 \Big\}\rho(x_1,x_2),
\end{eqnarray}
which was obtained from the master equation~(\ref{eq:master})
neglecting density-density correlations.
More precisely, the following approximation has been used
\begin{equation}
    \langle \hat{\Phi}^\dag (x_1)^2 \hat{\Phi}(x_1)
    \hat{\Phi}(x_2)\rangle\approx (\langle \hat{n}(x_1)\rangle
    - 1) \rho(x_1,x_2).
\end{equation}

Note that in the absence of the measurement, that is without the
last term in the right-hand side of Eq.~(\ref{eq:density-matrix}),
this equation is a simple generalization of the well-known
Gross-Pitaevskii equation (GPE) for the condensate wave-function.
This is easily seen by substituting the coherent state
 $\rho(x_1,x_2) \!=\!
\varphi^*(x_1) \varphi(x_2)$ into Eq.~(\ref{eq:density-matrix}). The
measurement, as can be seen from Eq.~(\ref{eq:density-matrix}),
results in decay of non-diagonal elements (coherence) of the
single-atom density matrix~\cite{caves}. This measurement-induced
decoherence prevents us from describing the gas in terms of the
condensate pure-state wave-function.

Using Eq.~(\ref{eq:density-matrix}) the evolution of single-atom
fluctuations can be derived as
\begin{eqnarray}
\label{eq:second-moments}
    \partial_t\langle \Delta x^2 \rangle &=& \frac{1}{m}\langle \{x,p\}\rangle, \\
    \partial_t\langle \Delta p^2 \rangle &=& -m\omega_0^2\langle \{x,p\} \rangle - 4 g_{\rm 1D} \langle n'(x) p \rangle  + \frac{2\kappa}{N^2},   \nonumber \\
    \partial_t \langle \{x,p\}\rangle &=& \frac{2}{m}\langle \Delta p^2 \rangle - 2 m\omega_0^2 \langle \Delta x^2 \rangle  - 4 g_{\rm 1D} \langle n'(x) x
    \rangle, \nonumber
\end{eqnarray}
where $\langle \{x,p\} \rangle$ denotes the anticommutator of $x$
and $p$, $n'(x)$ denotes the derivative of the density distribution
with respect to the coordinate. Here we restrict the consideration
to the states with $\langle x \rangle \!=\! \langle p \rangle \!=\!
0$. The system~(\ref{eq:second-moments}) is not closed since it
contains the terms proportional to $\langle n'(x) p \rangle$ and
$\langle n'(x) x \rangle$ that in general cannot be expressed via
the single-atom second moments only.

To render the system~(\ref{eq:second-moments}) closed we perform the
following rough approximation, which, as will be seen below, is
enough to grasp its qualitative behavior. First we note that without
interactions the continuous measurement of the CM coordinate does
not change the shape of the distribution function describing the
state of the atoms. This follows directly from the solution of the
Fokker-Planck equation (FPE) for the Wigner function $W(x,p)$ in the
case of the non-interacting gas
\begin{eqnarray}
\label{eq:FPE-nonint}
    \partial_t W(x,p) = \left(-\frac{p}{m}\partial_x + m\omega_0^2
    x \partial_p + \frac{\kappa}{N^2}\partial_p^2\right)W(x,p).
\end{eqnarray}
Solving this equation analytically with Gaussian initial condition
one easily finds that the measurement of BEC only changes the width
of the distribution preserving its Gaussian shape. We assume that in
the considered situation the distribution also remains approximately Gaussian
during the system evolution. This is certainly not true for the
strong interaction case~\cite{pitaevskii-rmp}. However, for weak
interactions this might be a rather good approximation. In this case
one obtains the following expressions for the sought averages
\begin{eqnarray}
    \langle n'(x) p \rangle &\approx& - \frac{N}{8\sqrt{\pi} \langle
    \Delta x^2
    \rangle^{3/2}}\langle \{x,p\} \rangle, \nonumber \\
    \langle n'(x) x \rangle &\approx& - \frac{N}{4\sqrt{\pi} \langle
    \Delta x^2 \rangle^{3/2}}\langle \Delta x^2 \rangle.
\end{eqnarray}
Substituting this result into the system~(\ref{eq:second-moments})
one obtains the following closed system of equations
\begin{eqnarray}
\label{eq:second-moments-1}
    \partial_t\langle \Delta x^2 \rangle &=& \frac{1}{m}\langle \{x,p\} \rangle, \nonumber \\
    \partial_t\langle \Delta p^2 \rangle &=& - m\Omega_{\rm eff}^2\langle \{x,p\} \rangle + \frac{2\kappa}{N^2},   \nonumber \\
    \partial_t\langle \{x,p\} \rangle &=& \frac{2}{m}\langle \Delta p^2 \rangle - 4 m \Omega_{\rm eff}^2 \langle \Delta x^2
    \rangle,
\end{eqnarray}
where the effective frequency $\Omega_{\rm eff}$ defined via
\begin{equation}
\label{eq:eff-frequency}
    \Omega_{\rm eff}^2 = \omega^2_0 -
    \frac{g_{\rm 1D} N}{2 \sqrt{\pi} m \langle \Delta x^2 \rangle^{3/2}}
\end{equation}
has been introduced.  The effective frequency is determined by the
size of the atomic localization domain, which is represented by
$\langle \Delta x^2 \rangle$. The set of nonlinear ordinary
equations~(\ref{eq:second-moments-1}) can easily be solved using one
of well established numerical techniques~\cite{numerical-recipes}.
We use predictor-corrector Adams scheme that is known to be well
suited for stiff problems.

\begin{figure}
  \begin{center}
    \includegraphics[width=0.5\textwidth]{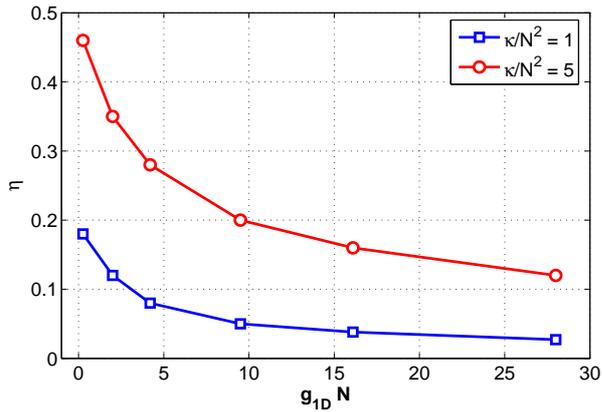}
    \caption{Relative spreading of the atomic cloud ($\eta$) as a function
    of the interaction strength for $\kappa/N^2 \!=\! 1$ (blue squares) and $\kappa/N^2 \!=\! 5$ (red circles).}
    \label{fig:eta-single-atom}
  \end{center}
\end{figure}

To characterize the effect of the CM position measurement on the gas
we introduce the so called relative spreading of the atoms $\eta$,
defined as
\begin{equation}
\label{eq:eta}
    \eta = \frac{\sqrt{\langle \Delta x^2 \rangle}_{\rm
    meas} - \sqrt{\langle \Delta x^2 \rangle}_{\rm
    no-meas}}{\sqrt{\langle \Delta x^2 \rangle}_{\rm meas}} .
\end{equation}
Here, the subscripts "meas" and "no-meas" are used to distinguish
the cases with and without the measurement, respectively. Taking for
the initial values of the fluctuations the results obtained from the
solution of the time-independent GPE we obtain the results shown in
Fig.~\ref{fig:eta-single-atom}.
In this plot the dependence of the relative spreading on the
interaction parameter $g_{\rm 1D} N$ is shown for two values of
measurement strength $\kappa/N^2$. Hereinafter, length is measured
in units of $a_0$, time in units of $T_0/2\pi$, where $T_0$ is the
period of oscillations in the harmonic potential. Thus, $g_{\rm 1D}
N$ is the dimensionless quantity.  The results of
Fig.~\ref{fig:eta-single-atom} correspond to the time instant
$t=T_0/4$, which is the time when initial momentum uncertainty
transforms to the position uncertainty. It is seen that being always
positive the relative spreading $\eta$ goes down for larger values
of the interaction constant. This means that the CM measurement
always increases the width of the atomic localization domain, but
for stronger atomic interactions the measurement-induced spreading
is slower. Thus, an interacting trapped gas appears to be more
stable against the measurement back-action noise than an ideal one,
at least in the beginning of the noise-governed evolution.

A simple explanation of this effect might be obtained on the basis
of Eq.~(\ref{eq:eff-frequency}). The effective frequency represents
a degree of atomic localization in an effective potential that is a
combination of the trap and mean-field potentials. According to
Eq.~(\ref{eq:eff-frequency}) the measurement induced spreading of
the cloud due to non-linear response also increases the effective
frequency, which corresponds to better localization. This mechanism partially
compensates for atomic delocalization due to the measurement
back action. As follows from Eq.~(\ref{eq:eff-frequency}) the effect
should be more pronounced for larger atom-atom coupling. However,
the validity of this approach is certainly limited by $g_{\rm 1D} N
\!<\! 2 \sqrt{\pi} m \omega_0^2 \langle \Delta x^2 \rangle^{3/2}$.

There is another interesting feature of the dynamics of an
interacting gas subjected to the CM position measurement. Note that
the CM of harmonically trapped gas is not coupled to the relative
motion~\cite{pitaevskii-rmp}. Thus, the measurement-induced
delocalization of the CM is increased regardless of atomic
interaction strength. The single atom distribution on the contrary
depends on the interaction and for strong interaction the
measurement-induced change of this quantity can be rather small.
This implies appearance of correlations corresponding to
\textit{bunching} of atoms and narrowing of the \textit{instant}
density profile compared with the initial unperturbed one.

\begin{figure}
  \begin{center}
    \includegraphics[width=0.4\textwidth]{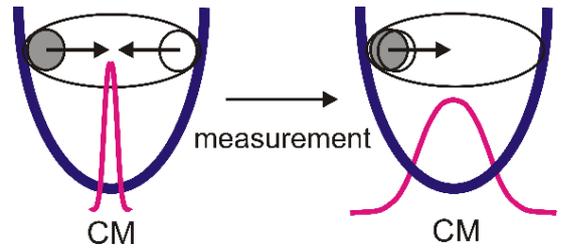}
    \caption{Two-atom illustration of the measurement-induced narrowing of the instant atomic density profile.
    The oval symbolizes the region of a single-atom localization that remains almost unchanged. The pulsed-shape distribution represents the
    CM position uncertainty that gradually increases due to the measurement back action. These two conditions can be
    simultaneously fulfilled if the atoms tend to bunch together.}
    \label{fig:sinchronization}
  \end{center}
\end{figure}

The easiest way to demonstrate this is to consider only two atoms. Assume
that initially each of them spread over the same region in the trap
(Fig.~\ref{fig:sinchronization}).
For the sake of simplicity let the atoms be spatially
anti-correlated (perform anti-phase oscillations). In this case, the
size of the atomic cloud is equal to the size of the single-atom
localization region, while the CM is strongly localized near the
center of the trap. The unconditioned measurement of the CM leads to
its delocalization, whereas the size of the atomic cloud, as
predicted by the theory described above, remains almost unchanged.
As illustrated in Fig.~\ref{fig:sinchronization} to ensure such
dynamics the atoms have to get closer to each other. This might
result in an instantaneous density profile that is narrower than the
initial ground-state one. In an experiment this effect manifests
itself in a size of a single (not ensemble averaged) resonance-image
or contrast-image snapshot that should become narrower after some
time of the system evolution.

\section{Numerical simulations}
\label{sec:numeric}

In this section we discuss the results of \textit{ab initio}
numerical simulations of the system. We use the numerical scheme
based on the positive P-representation~\cite{drummond-gardiner}. In
spite of known drawbacks~\cite{boundary-terms} of this approach it
is efficient for dynamical calculations restricted to relatively
short evolution times or small nonlinearities. In addition, this
method is relatively easy to implement.

\subsection{Phase-space representation}

To treat the problem numerically the continuous atomic distribution
is approximated by a lattice model. The space region occupied by the
atoms is divided into $M$ equal cells of length $\Delta x$: $x_i
\!=\! i \Delta x$, where $i \!=\! 1,\ldots, M$. For each cell $i$ we
define annihilation and creation operators
\begin{eqnarray}
\hat{a}_i &=& \int_{x_i-\Delta x/2}^{x_i+\Delta x/2} dx \hat{\Phi}(x)/\sqrt{\Delta x}, \nonumber\\
 \hat{a}_i^\dagger &=& \int_{x_i-\Delta x/2}^{x_i+\Delta x/2} dx\hat{\Phi}^\dagger(x)/\sqrt{\Delta
 x},
\end{eqnarray}
that obey commutation relations $[\hat{a}_i, \hat{a}_j] \!=\! 0$ and
 $[\hat{a}_i, \hat{a}^\dagger_j] \!=\! \delta_{ij}$. Then the
Hamiltonian~(\ref{eq:ham}) can be approximated by the following
Bose-Hubbard Hamiltonian
\begin{equation}
    \label{eq:ham-short}
    \hat{H} = \sum_{ij}\Upsilon_{ij} \hat{a}^{\dag}_i \hat{a}_j +
    \frac{g_{\rm 1D}}{\Delta x} \sum_i (\hat{a}_i^\dag)^2
    \hat{a}_i^2 .
\end{equation}
Here the matrix $\Upsilon$ accounts for the kinetic energy and potential energy in the trap, it
is defined as
\begin{equation}
    \label{eq:omega}
    \Upsilon_{ij} = \left(\frac{1}{\Delta x^2} +
    \frac{x_i^2}{2}\right)\delta_{ij} - \frac{1}{2\Delta x^2}
    \delta_{i+1 j} - \frac{1}{2\Delta x^2}
    \delta_{i-1 j}.
\end{equation}
The  discrete version of the non-Hamiltonian term of the master
equation~(\ref{eq:master}) is obtained in the similar way and need
not be explicitly written.

The total (many-atom) density operator is expanded using a positive
P-representation $P^{(+)}$
\begin{equation}
    \label{eq:ro-P-fun}
   \hat{\rho} = \int P^{(+)}(\textbf{a}) \hat{\Lambda}(\textbf{a})
   d^{2M}\!\boldsymbol{\alpha}\, d^{2M}\!\boldsymbol{\beta}
\end{equation}
using the following operator basis
\begin{equation} \label{eq:Lambda}
    \hat{\Lambda}(\textbf{a}) =  \frac{|\boldsymbol{\alpha}\rangle
    \langle \boldsymbol{\beta}^*|}{\langle \boldsymbol{\beta}^*|
    \boldsymbol{\alpha}\rangle}.
\end{equation}
Here, $\textbf{a} \!=\! (\boldsymbol{\alpha}, \boldsymbol{\beta})$,
where $\boldsymbol{\alpha} \!=\! \{\alpha_1, \ldots, \alpha_M\}$ and
$\boldsymbol{\beta} \!=\! \{\beta_1, \ldots, \beta_M\}$ are complex
vectors with components $\alpha_i \!=\! \alpha'_i \!+\! i
\alpha''_i$ and $\beta_i \!=\! \beta'_i \!+\! i \beta''_i$. The
$P^{(+)}$ representation is guaranteed to always produce
positive-definite diffusion, which is a necessary requirement for a
stochastic differential equation.

Substituting the expansion~(\ref{eq:ro-P-fun}) in the master
equation~(\ref{eq:master}) and using the standard operator identities
\cite{drummond-gardiner}
\begin{eqnarray}
    \label{eq:action-a-ro-1}
    \hat{\textbf{a}} \hat{\Lambda}(\textbf{a}) &=& \boldsymbol{\alpha} \hat{\Lambda}(\textbf{a}),
    \nonumber \\
    \hat{\textbf{a}}^\dag \hat{\Lambda}(\textbf{a}) &=& (\partial_{\boldsymbol{\alpha}} + \boldsymbol{\beta})\hat{\Lambda}(\textbf{a}),
    \nonumber \\
    \hat{\Lambda}(\textbf{a})\hat{\textbf{a}} &=& (\partial_{\boldsymbol{\beta}} + \boldsymbol{\alpha})\hat{\Lambda}(\textbf{a}),
    \nonumber \\
    \hat{\Lambda}(\textbf{a}) \hat{\textbf{a}}^\dag &=& \boldsymbol{\beta}
    \hat{\Lambda}(\textbf{a})
\end{eqnarray}
one can obtain the following FPE for the positive $P^{(+)}$-function
\begin{eqnarray}
\label{eq:P-eq-new} \frac{\partial}{\partial t} P^{(+)}(\textbf{a})
&=& \Big[-
\partial_\mu A_\mu + \frac{1}{2}\partial_\mu
\partial_\nu (\textbf{B}_{\rm int}\textbf{B}_{\rm int}^{T})_{\mu\nu} \\ \nonumber
&+& \frac{1}{2}\partial_\mu
\partial_\nu (\textbf{B}_{\rm meas}\textbf{B}_{\rm meas}^{T})_{\mu\nu}\Big]
P^{(+)}(\textbf{a}).
\end{eqnarray}
Here, $\partial_\mu$ denotes partial derivatives $\partial/\partial
\alpha_\mu$ if $\mu \leq M$ and $\partial/\partial \beta_{\mu -M}$
otherwise, $\mu$ and $\nu$ take values $\mu,\nu \!=\! 1,\ldots,2M$.
The elements of the drift vector $\textbf{A} \!=\!
\{A_1,\ldots,A_{M+1},\ldots \}$ are given by
\begin{eqnarray}
    \label{eq:drift}
    A_i &=& -i \Upsilon_{ij} \alpha_j - i \frac{g_{\rm 1D}}{\Delta x} \alpha_i^2 \beta_i - \frac{\kappa}{N^2} x_i^2 \alpha_i, \\
    \nonumber
    A_{i+M} &=& i \Upsilon_{ji} \beta_j + i \frac{g_{\rm 1D}}{\Delta x} \beta_i^2 \alpha_i - \frac{\kappa}{N^2} x_i^2 \beta_i.
\end{eqnarray}

The diffusion matrix can be divided into two parts corresponding
to different noise sources acting on the atoms. One of these sources may be attributed to the atom-atom interactions. It is described by the
diagonal matrix
\begin{equation}
    \label{eq:B-interaction}
    \textbf{B}_{\mathrm{int}} = \sqrt{\frac{g_{\rm 1D}}{\Delta x}} \mathrm{diag} \{(1-i)\alpha_1,\ldots, (1+i) \beta_1,\ldots\}.
\end{equation}
The other noise source is the measurement back action. This noise is
represented by the matrix $\textbf{B}_{\mathrm{meas}}$ with only one
(first) non-zero column. This matrix is written as
\begin{eqnarray}
\textbf{B}_{\mathrm{meas}} = -\frac{\sqrt{2\kappa}}{N} \begin{pmatrix} x_1 \alpha_1 & 0 & \ldots & 0  \\
\vdots & \vdots & \ddots & \vdots \\
x_1 \beta_1 & 0 & \ldots & 0 \\
\vdots & \vdots & \ddots & \vdots
\end{pmatrix}.
\end{eqnarray}
It is easy to check that the total diffusion matrix is given by
$\mathbf{D} = \textbf{B}_{\mathrm{int}}\textbf{B}_{\mathrm{int}}^{T}
+ \textbf{B}_{\mathrm{meas}}\textbf{B}_{\mathrm{meas}}^{T}$.
Assuming that the noise sources discussed above are represented by
\textit{independent} Wiener processes one can show that the
FPE~(\ref{eq:P-eq-new}) is equivalent to the set of It\^{o}
stochastic differential equations
\begin{eqnarray}
    \label{eq:stoch-eq-Ito-complex}
    d\textbf{a} &=& \textbf{A}(\textbf{a},t)dt +
    \textbf{B}_{\mathrm{int}}(\textbf{a},t) d\textbf{W}_{\mathrm{int}}(t) \nonumber \\
    &+&
    \textbf{B}_{\mathrm{meas}}(\textbf{a},t) d\textbf{W}_{\mathrm{meas}}(t).
\end{eqnarray}

For the numerical simulations it is instructive to obtain a set of
equations for real functions instead of
Eq.~(\ref{eq:stoch-eq-Ito-complex}). To do so we following, for
example, Ref.~\cite{operator-identities} decompose the drift vector
and the noise matrices into real and imaginary parts as $\textbf{A}
= \textbf{A}' \!+\! i\textbf{A}''$ and $\textbf{B} = \textbf{B}'
\!+\! i\textbf{B}''$. Since $\hat{\Lambda}(\mathbf{a})$ is an
analytic function the derivatives $\partial_{\alpha}$ and
$\partial_{\beta}$ can be evaluated in either real or imaginary
directions so that the resulting drift and diffusion terms can
always be made real. Taking this into account one can define the new
$4M$-dimensional real drift vector $\textbf{\b{A}} \!=\!
\{A'_1,\ldots,A'_{M+1},\ldots,A_1'',\ldots, A_{M+1}'',\ldots\}$ with
the elements
\begin{eqnarray}
    \label{eq:drift-real}
    A'_i &=& \Upsilon_{ij} \alpha''_j + \frac{g_{\rm 1D}}{\Delta x}(n''_i \alpha'_i + n'_i \alpha''_i) - \frac{\kappa}{N^2} x_i^2 \alpha'_i, \\
    \nonumber
    A''_i &=& -\Upsilon_{ij} \alpha'_j - \frac{g_{\rm 1D}}{\Delta x}(n'_i \alpha'_i - n''_i \alpha''_i) - \frac{\kappa}{N^2} x_i^2 \alpha''_i,
    \\ \nonumber
    A'_{i+M} &=& -\Upsilon_{ji} \beta''_j - \frac{g_{\rm 1D}}{\Delta x}(n''_i \beta'_i + n'_i \beta''_i)  - \frac{\kappa}{N^2} x_i^2
    \beta'_i, \\ \nonumber
     A''_{i+M} &=& \Upsilon_{ji} \beta'_j + \frac{g_{\rm 1D}}{\Delta x}(n'_i \beta'_i - n''_i \beta''_i)  - \frac{\kappa}{N^2} x_i^2
    \beta''_i.
\end{eqnarray}
Here, $n'_i \!=\! \alpha'_i \beta'_i \!-\! \alpha''_i \beta''_i$ and
$n''_i \!=\! \alpha'_i \beta''_i \!+\! \alpha''_i \beta'_i$ are real
and imaginary parts of the complex atom number $n_i \!=\! n'_i \!+\!
i n''_i$. The new stochastic matrices
$\textbf{\b{B}}_{\mathrm{int}}$ and $\textbf{\b{B}}_{\mathrm{meas}}$
are
\begin{eqnarray}
\label{eq:int-new}
    \textbf{\b{B}}_{\mathrm{int}} = \begin{pmatrix} \textbf{{\O}} & \textbf{B}_{\mathrm{int}}'  \\
    \textbf{{\O}} & \textbf{B}_{\mathrm{int}}'' \end{pmatrix},
\end{eqnarray}
with
\begin{eqnarray}
    \label{eq:B-new-interaction}
    \textbf{B}_{\mathrm{int}}' &=& \sqrt{\frac{g_{\rm 1D}}{\Delta x}} \mathrm{diag}\{\alpha'_1 + \alpha''_1,\ldots, \beta'_1 -
    \beta''_1,\ldots\}, \nonumber
    \\
    \textbf{B}_{\mathrm{int}}'' &=& \sqrt{\frac{g_{\rm 1D}}{\Delta x}} \mathrm{diag}\{-\alpha'_1 + \alpha''_1,\ldots,\beta'_1 + \beta''_1, \ldots\},
\end{eqnarray}
and
\begin{eqnarray}
\label{eq:meas-new}
    \textbf{\b{B}}_{\mathrm{meas}} = \begin{pmatrix} \textbf{{\O}} & \textbf{B}_{\mathrm{meas}}'  \\
    \textbf{{\O}} & \textbf{B}_{\mathrm{meas}}'' \end{pmatrix},
\end{eqnarray}
with
\begin{eqnarray}
    \textbf{B}'_{\mathrm{meas}} = \frac{\sqrt{2\kappa}}{N} \begin{pmatrix} - x_1 \alpha''_1 & 0 & \ldots & 0  \\
    \vdots & \vdots & \ddots & \vdots \\
    x_1 \beta''_1 & 0 & \ldots & 0 \\
    \vdots & \vdots & \ddots & \vdots
\end{pmatrix}
\end{eqnarray}
and
\begin{eqnarray}
    \textbf{B}''_{\mathrm{meas}} = \frac{\sqrt{2\kappa}}{N} \begin{pmatrix} x_1 \alpha'_1 & 0 & \ldots & 0  \\
    \vdots & \vdots & \ddots & \vdots \\
    -x_1 \beta'_1 & 0 & \ldots & 0 \\
    \vdots & \vdots & \ddots & \vdots
\end{pmatrix}.
\end{eqnarray}
The matrix $\textbf{{\O}}$ in Eqs.~(\ref{eq:int-new}) and (\ref{eq:meas-new}) denotes the $2M\times 2M$ matrix with zero elements.

The SDE~(\ref{eq:stoch-eq-Ito-complex}) is then cast into the
following form containing new real $4M$-dimensional Wiener noise
vectors $\textbf{\b{W}}_{\mathrm{int}}$ and
$\textbf{\b{W}}_{\mathrm{meas}}$:
\begin{eqnarray}
    \label{eq:stoch-eq-Ito}
    d\textbf{\b{a}} &=& \textbf{\b{A}}(\textbf{\b{a}},t)dt +
    \textbf{\b{B}}_{\mathrm{int}}(\textbf{\b{a}},t) d\textbf{\b{W}}_{\mathrm{int}}(t) \nonumber \\
    &+&
    \textbf{\b{B}}_{\mathrm{meas}}(\textbf{\b{a}},t) d\textbf{\b{W}}_{\mathrm{meas}}(t).
\end{eqnarray}
The 4M dimensional real vector $\textbf{\b{a}}$ is formed of real
and imaginary parts of $\boldsymbol{\alpha}$ and
$\boldsymbol{\beta}$. The elements of noise vectors
$d\textbf{\b{W}}_{\mathrm{int}}$ and
$d\textbf{\b{W}}_{\mathrm{meas}}$ with the elements denoted by
$d\b{W}^{(\mathrm{int})}_i$ and $d\b{W}^{(\mathrm{meas})}_i$ obey
the following properties
\begin{eqnarray}
\label{eq:wiener-properties}
    \langle d\b{W}^{(\mathrm{meas})}_i \rangle = \langle d\b{W}^{(\mathrm{int})}_i \rangle = 0, \nonumber \\
    \langle d\b{W}^{(\mathrm{int})}_i d\b{W}^{(\mathrm{int})}_j \rangle = \delta_{ij}dt, \nonumber \\
    \langle d\b{W}^{(\mathrm{meas})}_i d\b{W}^{(\mathrm{meas})}_j \rangle = \delta_{ij}dt, \nonumber \\
    \langle d\b{W}^{(\mathrm{int})}_i d\b{W}^{(\mathrm{meas})}_j \rangle
    = 0, \,\,\, \forall \,\,\, i,j.
\end{eqnarray}
Note that the measurement noise matrix
$\textbf{\b{B}}_{\mathrm{meas}}$ consists of a single non-zero
column. This means that all the modes (lattice cells) of the trapped
gas are affected by the same measurement-induced noise. This is
expectable since the considered noise acts on the collective (CM)
degree of freedom of the system. The noise matrix
$\textbf{\b{B}}_{\mathrm{int}}$ that originates from the atomic
interactions is diagonal, which means that each space point of the
gas is driven by its individual Wiener noise. Noise sources acting
on different coordinates of the gas are statistically independent.

\subsection{Results of numerical simulation}

The numerical solution of SDE~(\ref{eq:stoch-eq-Ito}) has been found
using semi-implicit method discussed in
Ref.~\cite{drummond-mortimer}. This approach is generally quite
stable when applied to nonlinear and/or stiff problems. However the
problem under consideration seems to have intrinsic instability
which manifests itself during the simulations regardless of the
stability of used numerical technique. Such an instability, as
discussed in Ref.~\cite{boundary-terms}, is a result of incorrectly
ignored boundary term during the derivation of FPE for the positive
P-function.

The conditions when the problem becomes unstable can be grasped
looking at the terms of SDE~(\ref{eq:stoch-eq-Ito-complex})
proportional to the interaction constant. As soon as the system
evolves to a quantum state with $\boldsymbol{\alpha}^* \neq
\boldsymbol{\beta}$ these terms may acquire a real part responsible
for exponential growth. Thus considerable (in some sense) deviation
of the $\boldsymbol{\beta}$ from $\boldsymbol{\alpha}^*$ indicates
the limits of applicability of the approach.

In a series of numerical experiments performed for different values
of the interaction constant the evolution time has been determined
during that the dynamics demonstrates no sign of "exploding"
trajectories. This "secure" time interval is about quarter of the
trap oscillation period. Thus the numerical simulation results
presented below correspond to $T_0/4$. The spatial discretization
used for the calculation is $\Delta x \!=\! 0.33$, the time step is
$\Delta t \!=\! 10^{-4}$. Further decrease of these values does not
change the appearance of the plots presented below. Thus the
discretization error can be estimated as being of the order of the
line thickness. The number of the stochastic trajectories equals to
$20000$. The estimated sampling error in this case is about 0.03\%
which is also below the thickness of the plot lines.

For the initial state of the system it is convenient to take BEC broken-symmetry coherent state. This state for the lattice model reads
\begin{equation}
\label{eq:initial-state}
    | \Psi \rangle = | \varphi_{x_1} \rangle
    \otimes \ldots \otimes | \varphi_{x_M} \rangle,
\end{equation}
where $\varphi_{x_i}$ is the value of the solution of the
time-independent GPE in the space point $x_i$. The state gives the
following initial values for the phase-space variables
\begin{eqnarray}
\label{eq:initial-state-1}
    \alpha'_i(0) &=& \rm{Re}(\varphi_{x_i}),
    \,\,\,
    \beta'_i(0) = \alpha'_i(0), \nonumber \\
    \alpha''_i(0) &=& \rm{Im}(\varphi_{x_i}), \,\,\, \beta''_i(0) =
    \alpha''_i(0).
\end{eqnarray}

\begin{figure}
  \begin{center}
    \includegraphics[width=0.5\textwidth]{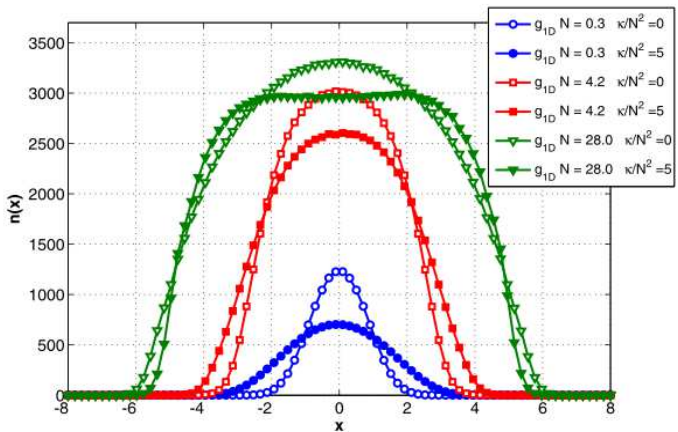}
    \caption{Atom density profiles for different coupling constants.
    The curves shown by empty and filled symbols correspond
    to the situations with and without the measurement, respectively.}
    \label{fig:density}
  \end{center}
\end{figure}

The results of numerical calculations of the atom density profile for different values of the
interaction strength $g_{\rm 1D}N$ are shown in Fig.~\ref{fig:density}. The curves shown by empty and filled
symbols correspond to the situations with and without the measurement, respectively.
For small interaction strength the density profiles corresponding to these cases differ substantially (circles).
The difference becomes less essential with the increase of $g_{\rm 1D}N$ (squares).
For yet stronger interaction the effect of the measurement becomes practically negligible, compare curves plotted with filled and
empty triangles.

\begin{figure}
  \begin{center}
    \includegraphics[width=0.5\textwidth]{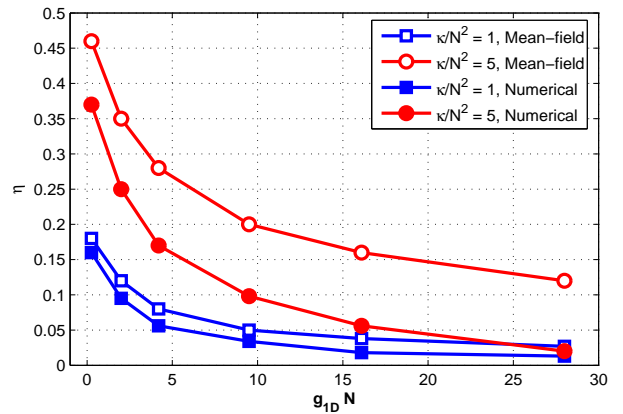}
    \caption{Relative spreading of the cloud ($\eta$) as a function of the interaction strength.
    Filled and empty symbols show the results of the numerical simulation and the mean-field approximation, respectively. Squares (blue) and circles (red)
    correspond to different values of the measurement resolution.}
    \label{fig:eta-comparison}
  \end{center}
\end{figure}

Figure~\ref{fig:eta-comparison} compares the relative spreading of
the atoms $\eta$~(\ref{eq:eta}) as a function of the interaction strength for numerical
simulation (filled symbols) and the mean-field approach of Sec.~\ref{sec:single-atom} (empty
symbols). Curves shown by squares (blue) and circles (red) correspond to
different values of $\kappa/N^2$. It is seen that both methods predict similar
qualitative behavior of the relative atomic spreading. That is the numerical simulations confirm predicted earlier
decrease of spreading $\eta$ with increasing of atom-atom interaction strength.
However, some quantitative discrepancy is observed, which is found
to be more pronounced for larger $g_{\rm 1D}N$. This is not surprising since in
deriving the mean-field approach a couple of not-well-justified assumptions have been made. One of them is the assumption that the
density-density type correlations are small. This is exactly the case for the chosen initial state~(\ref{eq:initial-state}) but may become wrong
after some time of the system evolution. The other assumption that is certainly violated for a strongly interacting gas is the gaussian
density profile approximation. These poor approximations of the mean-field approach result in quantitative difference between the results of the two
methods. Nevertheless, the essential feature of interacting gas dynamics subjected to the CM position measurement can be grasped within the mean-field
approximation as derived above.

The qualitative arguments presented at the end of Sec.~\ref{sec:single-atom} indicate the possibility of generating
bunched states. These states are also characterized by squeezed compared with the unperturbed BEC ground state density profile.
Clearly the discussed bunching effect should manifest itself in the second-order correlation function, which can be defined as
\begin{equation}
\label{eq:second-order-corr} g_2(x) = \frac{\langle
\hat{\Phi}^{\dag}(0) \hat{\Phi}^{\dag}(x) \hat{\Phi}(x)
\hat{\Phi}(0)\rangle}{\langle \hat{\Phi}^{\dag}(0) \hat{\Phi}(0)
\rangle \langle \hat{\Phi}^{\dag}(x) \hat{\Phi}(x) \rangle}.
\end{equation}
This quantity characterizes density-density correlations between the trap center and
the point with the coordinate $x$. In case of independent densities in these points $g_2(x) \!=\! 1$, increased (decreased) likelihood
to detect atoms separated by $x$ means $g_2(x) > 1$ ($g_2(x)<1$).
\begin{figure}
  \begin{center}
    \includegraphics[width=0.45\textwidth]{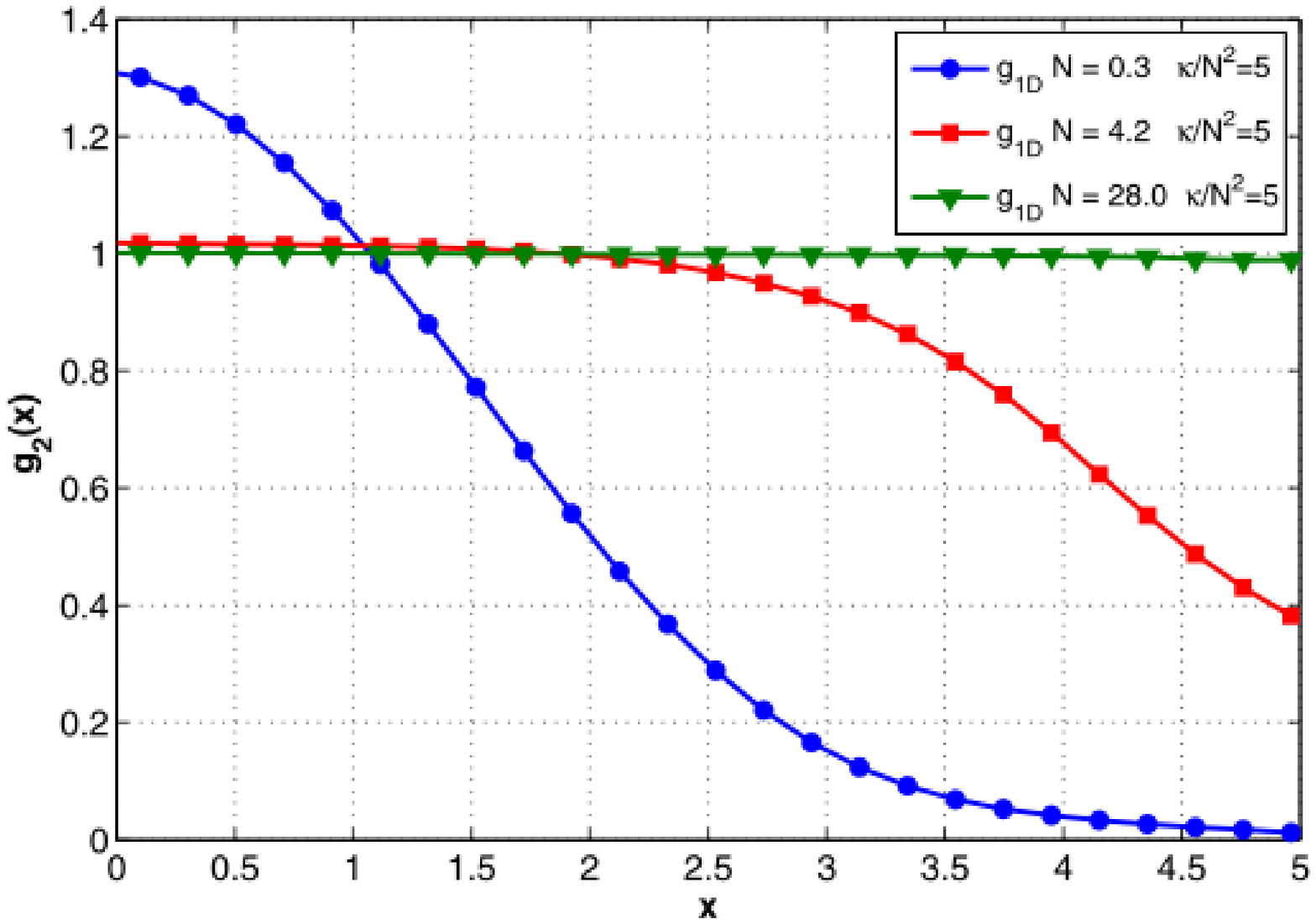}
    \caption{Second-order correlation functions for $\kappa/N^2 \!=\!
    5$} and various values of the atom-atom interaction strength $g_{\rm 1D}N$.
    \label{fig:g2}
  \end{center}
\end{figure}

Figure~\ref{fig:g2} shows these density-density correlations for
$\kappa/N^2 \!=\!5$ and different values of atom-atom interactions.
It is seen that for small coupling strength $g_{\rm 1D}N$ the
measurement of the collective coordinate leads to the well observed
bunching of the atoms (the curve shown with circles). This indicates
that inside the space region occupied by the atoms they are
distributed non-uniformly, that is the atoms are grouped in a bunch
with the size smaller than that of the occupation region. This is a
natural result taking into account that the measurement back-action
noise acts on the collective observable resulting in the overall
delocalization of the atoms while keeping the initial
\textit{instant} density distribution.

For larger coupling strength, the curve shown with squares in
Fig.~\ref{fig:g2}, the value of the second-order correlation
function $g_2(x)$ is smaller for the small separations, $x
\rightarrow 0$, than in the previous case. This indicates the
smaller degree of bunching that is the result of increased atom-atom
repulsion. For yet stronger repulsive interaction (the curve shown
with triangles in Fig.~\ref{fig:g2}) the value of the second-order
correlation function $g_2(0)$ is only slightly larger than one
(cannot be seen on the plot). Thus, the atom density in different
locations becomes independent. This is in contrast with the closed
strongly interacting gas which exhibits
anti-bunching~\cite{gas-anti-bunching}.

It is worth noting that for weak (blue curve) and moderate (red
curve) interactions the second-order correlation function does not
seem to approach unity even for large separations. We believe that
this is the result of collective character of considered back
action. That means that essentially all the atoms are simultaneously
subjected to the same back-action noise, which results in the
observed in Fig.~\ref{fig:g2} long-range correlations.

However we consider a finite system where the limit of very large
separations is somewhat ambiguous. At least in numerical
calculations of the second-order correlation function we have to
restrict the separations to be smaller than the size of the atomic
cloud, otherwise the use of definition~(\ref{eq:second-order-corr})
becomes impractical due to roundoff errors. To summarize the
discussion, the presented numerical results only indicate the
formation of the long-range correlations. The detailed description
of the second-order correlation function in the limit of large
separations can be obtained using a properly defined thermodynamic
limit.

\section{Summary}
\label{sec:summary}

The CM position measurement of trapped ultra-cold gases can become
an important ingredient of various technologies based on ultra-cold
atoms. The effect of such a measurement performed on quasi 1D
harmonically trapped interacting Bose-gas has been studied in this
article. The CM measurement back action disturbs the momentum, which
due to the oscillations in the trap results in atomic
delocalization. It has been shown that the interaction-induced
nonlinearity can to some extent resist this delocalization
decreasing the rate of growth of the width of the average density
profile. In other words, the atomic cloud size grows slowly for the
stronger interacting gas. This result has first been obtained using
a semi-analytical mean-field approach with certain density-density
type correlations ignored. Then the same conclusion has been
obtained on a more rigorous basis performing numerical simulations
based on positive P-representation of the many-atom density
operator. In these simulations the atomic correlations have been
taken into account, but the continuous density distribution has been
approximated by a lattice model.

Numerical simulations show that the averaged atom density profile or
atomic cloud size remains almost unchanged for strongly interacting
gas during the first quarter of the oscillation period. The
numerical approach has allowed to calculate second-order correlation
functions in the presence of the CM position measurement. The value
of this correlation function for small atomic separation has been
found to be greater than one for weakly and even strongly
interacting gases, which is the atomic bunching.

In addition, some preliminary conclusions can be made about the
width of the instant atomic density profile. As discussed above, the
average density profile of strongly interacting gas is almost
unchanged in the beginning of the evolution. However, the CM
position uncertainty being decoupled from the internal degrees of
freedom constantly grows due to measurement back action noise. This
is possible if the instant profile shrinks below its initial value.
The latter conclusion is to be verified by direct calculations,
which will be given elsewhere.

\section{Acknowledgements}
The authors thank V.V. Kozlov for valuable discussions. One of the
authors (DAI) is grateful to S. Wallentowitz for the introduction to
the theory of open many-atom systems. DAI also acknowledge Saint
Petersburg government for the research grant. The authors
acknowledge Saint Petersburg State University for a research grant.
MSS acknowledges financial support from "Dynasty" foundation.


\begin{thebibliography}{20}
\bibitem{atomic-clock} J.~Dunningham, K.~Burnet, and W.~D.~Phillips, Phil. Trans. R. Soc. A \textbf{363}, 2165 (2005).
\bibitem{drummond-leuchs} T.~Vaughan, P.~Drummond, and G.~Leuchs, Phys. Rev. A \textbf{75}, 033617 (2007).
\bibitem{casimir-polder} M.~Antezza, L.~P.~Pitaevskii, and S.~Stringari, Phys. Rev. A \textbf{70}, 053619 (2004); D. M. Harber, \textit{et al.},
Phys. Rev. A \textbf{72}, 033610 (2005).
\bibitem{QND} V.~B.~Braginsky and F.~Ya.~Khalili, Quantum measurement: Cambridge University Press, 1992.
\bibitem{brennecke} F.~Brennecke, \textit{et al.}, Nature
\textbf{450}, 268 (2007).
\bibitem{colombe} Y.~Colombe, \textit{et al.}, Nature \textbf{450},
272 (2007).
\bibitem{gupta} S.~Gupta, \textit{et al.}, Phys. Rev. Lett.
\textbf{99}, 213601 (2007).
\bibitem{murch} W.~W.~Murch, \textit{et al.}, Nature Phys.
\textbf{4}, 561 (2008).
\bibitem{wiseman-bec} H.~M.~Wiseman and L.~K.~Thomsen, Phys. Rev. Lett. \textbf{86}, 12 (2001).
\bibitem{wallentowitz-prl} D.~Ivanov and S.~Wallentowitz, Phys. Rev. Lett. \textbf{93}, 260603 (2004).
\bibitem{pitaevskii-rmp} F.~Dalfovo, \textit{et al.}, Rev. Mod. Phys. \textbf{71}, 463 (1999).
\bibitem{ng} G.~S.~Ng, \textit{et al.}, New Journal of Physics \textbf{11}, 073045 (2009).
\bibitem{petrucione-book} H.-P.~Breuer and F.~Petruccione, The Theory of Open Quantum Sysetms: Oxford University Press, 2002.
\bibitem{drummond-gardiner} P.~D.~Drummond and C.~W.~Gardiner, J. Phys. A: Math. Gen. \textbf{13}, 2353 (1980).
\bibitem{boundary-terms} A.~Gilchrist, C.~W.~Gardiner, and P.~D.~Drummond, Phys. Rev. A \textbf{55}, 3014 (1997).
\bibitem{coupling-const} M. Olshanii, Phys. Rev. Lett. {\bf 81}, 938 (1998).
\bibitem{mekhov} I.~B.~Mekhov, C.~Maschler, and H.~Ritsch, Nature Phys. \textbf{3}, 319 (2007);
I.~B.~Mekhov and H.~Ritsch, Phys. Rev. Lett. \textbf{102}, 020403 (2009).
\bibitem{reichel-nature} Y.~Colombe, \textit{et al.}, Nature \textbf{450}, 272
(2007).
\bibitem{hutchinson} N.~I.~Kaliteevsky, Wave Optics: Vysshaya Shkola, 1978 (in Russian).
\bibitem{caves} C.~M.~Caves and G.~J.~Milburn, Phys. Rev. A \textbf{36}, 5543 (1987).
\bibitem{numerical-recipes} W.~H.~Press, \textit{et al.}, Numerical Recipes in C: The Art of Scientific Computing:
Cambridge University Press, 1997.
\bibitem{operator-identities} P.~Deuar and P.D.~Drummond, Phys. Rev. A {\bf 66}, 033812 (2002).
\bibitem{drummond-mortimer} P.D.~Drummond and I.K.~Mortimer, J. Comp.
Phys. {\bf 93}, 144 (1991).
\bibitem{gas-anti-bunching}  M.~Naraschewski, R.J.~Glauber, Phys. Rev. A \textbf{59}, 4595 (1999).
\end{thebibliography}
\end{document}